\documentstyle[aps,epsfig]{revtex}
\begin{document}
\draft
\title{Modelling sublimation and atomic layer epitaxy in the presence of 
	competing surface reconstructions}
\author{M. Ahr\cite{myself} $^{1}$ \and M. Biehl$^{1,2}$}
\address{$^{1}$Institut f\"{u}r Theoretische Physik und Astrophysik, 
	Julius-Maximilians-Universit\"{a}t W\"{u}rzburg, Am Hubland, 97074 
	W\"{u}rzburg, Germany\\ 
	$^{2}$ Sonderforschungsbereich 410, Universit\"{a}t W\"{u}rzburg}
\date{\today}
\maketitle

\begin{abstract}
We present a solid-on-solid model of a binary {\it AB} compound, where atoms 
of type {\it A} in the topmost layer interact via anisotropic  
interactions different from those inside the bulk. Depending on temperature 
and particle flux, this model displays surface reconstructions similar 
to those of (001) surfaces of II-VI semiconductors. 
We show, that our model qualitatively reproduces many of the 
characteristic features of these materials which have been observed during 
sublimation and atomic layer epitaxy.
We predict some previously unknown effects which might be observed 
experimentally.  
\end{abstract}

Apart from a few exceptions (eg. \cite{ibajjv98}), theoretical models of crystal surfaces 
in {\em nonequilibrium} situations have not
taken into account the effects of surface reconstructions, so far. Consequently, 
a theoretical understanding of the interplay of this phenomenon with the dynamics of 
growing or sublimating surfaces has not yet been achieved. In this paper we will 
address this important question in the context of a compound material that displays 
a competition of different vacancy structures in the terminating layer. 

Within the last years, there have been intensive experimental studies 
concerning the surfaces of II-VI semiconductors, see eg. \cite{pic26,ct97}
for an overview. These studies have revealed the existence of 
a fairly small number of surface reconstructions, which makes them promising 
candidates for a theoretical modelling. 
In the following, we will refer mostly to experimental results for CdTe(001) 
[2-9]. Under vacuum conditions, the CdTe(001) surface is metal terminated. 
The Cd atoms arrange in a vacancy structure, where half of the potential 
Cd sites are empty. At low temperatures, a $c(2\times2)$ reconstruction
is observed, in which Cd atoms arrange in a checkerboard like configuration. 
Often a contribution of $(2\times1)$ arrangement in rows along the (110) direction
is observed. 
At a temperature of $\approx 300^{0}C$, a  
reordering occurs, above which a $(2\times1)$ ordering dominates
on a sublimating surface. 
An additional Cd flux stabilizes the $c(2\times2)$ ordering even 
at high temperatures. Under a Te flux, the surface is Te-terminated, with a
$(2\times1)$ reconstruction. At small Te fluxes, the Te-coverage is 1, at 
low temperatures and high Te fluxes a higher coverage of $1.5$ is observed
\cite{ct97,tdbev94,ntss00,dbt96}.
In this Letter, we present a 2+1-dimensional solid-on-solid model of 
a binary compound material to get insight into the following problems: 
(1) Is the claim \cite{ntss00} of an effective thermal equilibrium of the 
surface layer justified under the non-equilibrium conditions of sublimation?  
(2) What is the effect of an external particle flux on the surface reconstruction?
(3) How do reconstructions influence techniques of crystal growth like atomic 
layer epitaxy (ALE)? 

In the terminating layer of a CdTe(001) surface, the potential Cd sites form a 
regular square lattice, where the simultaneous occupation of 
nearest neighbour (NN) sites in 
the $[0\overline{1}0]$-direction (denoted as y in the following) is forbidden by 
electron counting rules \cite{p89}. In our model, there is an attractive interaction between 
nearest neighbours in $[110]$(x)-direction (coupling $\varepsilon_{x}$) and diagonal 
neighbours (coupling $\varepsilon_{d}$).
Measuring energy in units of $|\varepsilon_{d}|$, we set $\varepsilon_{d} = -1$. 
A model, where the underlying crystal was assumed to be fixed, has been studied elsewhere
\cite{baksv00}. If the energy 
difference between a perfect $c(2\times2)$ arrangement and a perfect 
$(2\times1)$ arrangement is small and positive 
($-2 < \varepsilon_{x} \leq -1.9$), this model explains the $c(2\times2)$-$(2\times1)$ 
reordering as an accompanying effect of an order-disorder transition in thermal equilibrium. 
At low 
temperature $T$, the system is in an {\em ordered} phase, where $c(2\times2)$  
dominates, above the transition it is in a {\em globally} 
disordered phase, where the $(2\times1)$ rows dominate the {\em local} 
environment of a particle.

\begin{figure}
\begin{center}
\epsfig{file=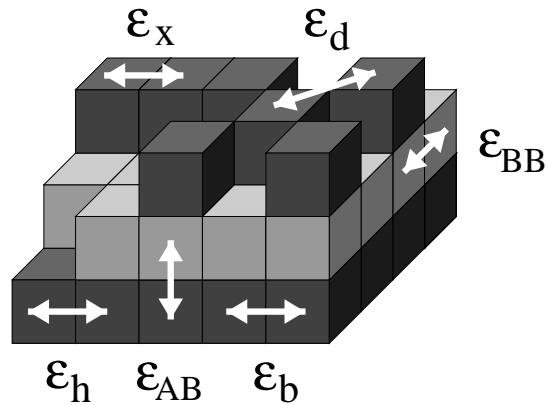,angle=270,width=0.4\columnwidth,clip=}
\end{center}
\caption{Sketch of a surface showing the interactions between the particles. {\it A} particles 
are shown as dark cubes, {\it B} particles in light gray.}
\label{fig3}
\end{figure}

In this paper, we  extend this planar lattice gas to a model of a 
three-dimensional crystal. 
Currently no values of diffusion energy barriers from {\em ab initio} 
or semi-empirical calculations are available, and the existing experimental 
data are insufficient for a parameter fit. Therefore, a realistic modelling of CdTe 
is beyond the scope of this work, which aims at an understanding of 
fundamental properties of nonequilibrium crystal surfaces. 
Instead, we consider a model of a compound ``{\it AB}'' with a cubic lattice
and a comparatively simple potential energy surface
which, although not being quantitative, reproduces essential features 
observed in experiments on CdTe(001). 
Here, the {\it A} ({\it B}) atoms correspond to Cd (Te).
Within the framework of a solid-on-solid model, we describe the crystal 
by a two-dimensional array $\{h_{xy}\}_{x, y = 1}^{N}$ of integers which 
denote the height of a column of atoms. We use periodic boundary 
conditions.   
The interactions between the particles are described by phenomenological 
parameters which are supposed to include the effects of surface strain and 
other interactions. To model the layered structure of 
the CdTe crystal, we assign the odd heights to {\it A} particles, and the even 
heights to {\it B} particles. 
Inside the bulk of the crystal, there is an attractive interaction 
between the particles and their NNs which is isotropic in directions 
parallel to the surface. 
While there is no difference in the interaction of {\it B} particles between the bulk
and the surface, {\it A} particles on the surface interact with the anisotropic 
interactions $\varepsilon_{x}$ and $\varepsilon_{d}$. 
The other parameters of the model are defined in Figure \ref{fig3}.
Thus, the Hamiltonian of the crystal is
\begin{equation}
H = \varepsilon_{AB} n_{AB} + \varepsilon_{BB} n_{BB} + \varepsilon_{b} 
n_{AA}^{b} + \varepsilon_{h} n_{AA}^{h}+ \varepsilon_{d} n_{AA}^{d} + 
\varepsilon_{x} n_{AA}^{x}  
\end{equation}
Here, $n_{AB}$ is the number of {\it AB} bonds between the layers, 
$n_{BB}$ is the number of NN pairs of {\it B} atoms, $n_{AA}^{b}$
the number of {\it A} NN pairs inside the bulk, $n_{AA}^{h}$ the 
number of surface {\it A} atoms next to a higher column, $n_{AA}^{d}$ ($n_{AA}^{x}$)
the number of bindings of surface {\it A} atoms to diagonal neighbours 
(NNs in x-direction) at the same height. The stay of particles 
in the wrong sublattice and the formation of NN pairs of 
surface {\it A} atoms in y-direction is forbidden. We allow diffusion both 
to nearest neighbour sites and to diagonal neighbours. Additionally, 
we permit the diffusion of an {\it AB} pair with the {\it B} atom on top of the 
{\it A} atom, if the diffusion of the {\it B} atom alone would end in the wrong 
sublattice. This process is required to preserve the ergodicity of the 
system. 
We simulate a Kawasaki dynamics, i.e. the rate of diffusion events is 
$\nu_{0} \exp(- (B_{0} + \Delta H)/T)$ if $\Delta H > 0$ and 
$\nu_{0} \exp(-B_{0}/T)$ else. 
The desorption of an {\it A} ({\it B}) particle requires an additional activation 
energy $E_{A}$ ($E_{B}$). This is done using rejection free Monte-Carlo techniques 
employing a binary search tree \cite{nb99}.
Particles from an external source arrive at each lattice site with equal 
probability. If the adsorption of the particle at this site would lead to a 
forbidden state, it is destroyed. 
 
This dynamics does not depend on the parameter 
$\varepsilon_{AB}$, so it needs not to be 
specified, and $\varepsilon_{BB}$ and $\varepsilon_{b}$ enter only via the sum $\varepsilon' := \varepsilon_{BB} + 
\varepsilon_{b}$. The constant prefactor $1/t_{0} := \nu_{0} \exp(-B_{0}/T)$
sets the timescale of the model and needs not to be specified, if we measure
time in units of $t_{0}$.  

In the following, we will use the parameter set 
$\varepsilon' = \varepsilon_{d} = -1$, $\varepsilon_{h} = 0$, 
$\varepsilon_{x} = -1.97$, $E_{A} = E_{B} = 1.5$ 
(in units of $|\varepsilon_{d}|$). The fact that the difference in the 
surface energies between perfectly 
$c(2\times2)$ and $(2\times1)$ reconstructed surfaces is small is consistent 
with DFT calculations \cite{gffh99}.
However, precise values of binding energies as a 
quantitative input to our model are not available. 
We have verified, that the behaviour of our model remains qualitatively the same for 
a wide range of parameters, as long as the basic property of a small energy difference between 
the reconstructions is preserved. 

To characterize the surface reconstruction quantitatively, we measure 
the {\it A}-coverage $\theta_{A} = n_{A}^{surf.}/N^{2}$, and the normalized 
correlations $C_{AA}^{d} :=  n_{AA}^{d}/(4 n_{A}^{surf.})$ and 
$C_{AA}^{x} := n_{AA}^{x}/(2 n_{A}^{surf.})$, which quantify the fraction 
of {\it A}-atoms on the surface which are incorporated in locally 
$c(2\times2)$ or $(2\times1)$ reconstructed areas, respectively. 
Additionally, we calculate LEED intensities to 
model diffraction experiments. Neglecting multiple scattering, this is 
done by calculating the fourier transform of the array 
$\{ \exp(i \, \Delta k \, h_{xy})\}_{x,y = 1}^{N}$, where equal diffraction 
intensities for {\it A} and {\it B} atoms have been assumed. The antiphase condition 
between different layers is $\Delta k = \pi/2$. 

\begin{figure}
\begin{center}
\epsfig{file=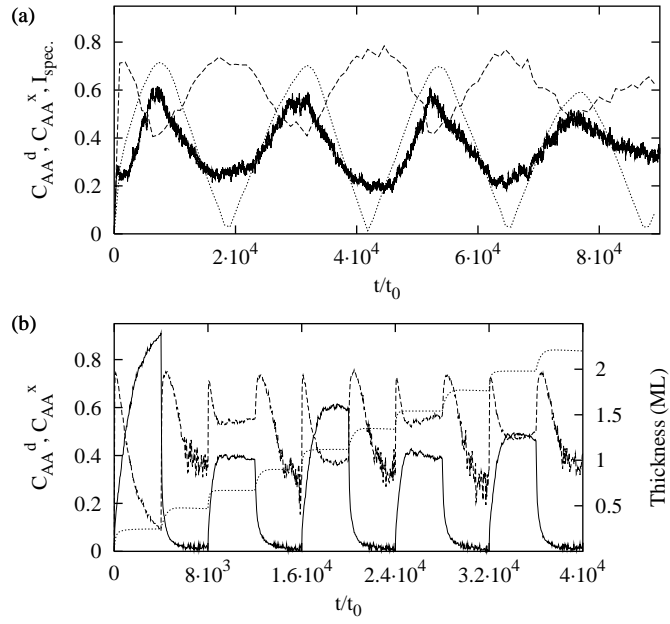,width=0.5\columnwidth,clip=}
\end{center}
\caption{Panel (a): Sublimation starting from a flat {\it B}-terminated surface 
at $T = 0.5$ and a system size $N = 128$. Solid lines: $C_{AA}^{d}$, dashed lines: 
$C_{AA}^{x}$, dotted: $I_{spec.}$(arbitrary units).  
For clarity of plotting, the last two curves have been smoothed; $C_{AA}^{d}$ shows the natural fluctuations. Each oszillation period corresponds 
to the desorption of one ML. Panel(b): Results of ALE simulations: $C_{AA}^{d}$(solid), 
$C_{AA}^{x}$(dashed) and the thickness of the deposited film (dotted) 
at $T = 0.3$, $t_{cykl} = 4\cdot10^3 t_{0}$ 
and a particle flux of 20 ML/cycle. The system size was $N = 256$. 
}
\label{fig1}
\end{figure}

Figure \ref{fig1}a shows the evolution of an initially {\it B}-terminated, flat 
surface under vacuum at a temperature $T = 0.5$. Initially, $\theta_{A}$ increases 
from zero to an asymptotic value close to 0.5. We find, 
that this increase can be fitted with an exponential relaxation, 
$\theta_{A} = \theta_{A}^{\infty} \left( 1 - \exp(-t/\tau)\right)$. Here, $\tau$ is 
the time constant of the decay of the {\it B}-terminated surface. We find, that 
the temperature dependence of the corresponding desorption probability 
$p \propto 1/\tau$ follows an Arrhenius law, 
$p \propto \exp(- E_{act.}/T)$, with an activation energy 
$E_{act.} = B_{0} + 3.0 \pm 0.1$ After this onset, {\it A} and {\it B} evaporate 
stoichiometrically, and sublimation proceeds in layer-by-layer mode, 
which reflects in the oszillations of the specular LEED intensity $I_{spec}$ 
in antiphase (dotted curve in Figure \ref{fig1}a). 
The dependence of the sublimation rate on $T$ 
follows an Arrhenius law with a higher activation energy  $B_{0} + 6.0 \pm 0.2$. 
Thus, our model reproduces the observation
of Cd-terminated CdTe surfaces under vacuum. Studies of surface lifetimes
\cite{tdbev94} and QMS measurements of evaporation rates \cite{nskts00} 
have also revealed significantly smaller activation energies for the decay of the 
Te-terminated surface ($\approx 0.7 eV$) than for congruent 
evaporation ($1.94 eV$ in layer-by-layer mode). We note however, that in our 
model the {\em microscopical} energy barriers for desorption of {\it A} and {\it B} 
are equal. Therefore the difference in the {\em macroscopic} activation 
energies is solely due to the stabilizing effect of the surface reconstruction.  

Surprisingly, $C_{AA}^{d}$ and $C_{AA}^{x}$ oszillate during 
sublimation, which proceeds layer by layer. Each time a complete 
layer has desorbed and the surface is atomically flat (maxima of
$I_{spec}$) the $c(2\times2)$ 
reconstructed fraction of the surface is maximal. On the contrary, 
a rough surface with a large number of islands (minimal $I_{spec}$) 
seems to prefer the $(2\times1)$ reconstruction. 
This can be understood from the fact that the attractive lattice gas 
interactions are present only between particles in the same layer. 
Thus, the island edges impose open boundary conditions to the lattice gas of 
{\it A} atoms on the island. In contrast to a $c(2\times2)$ reconstructed domain, a
$(2\times1)$ terminated island can {\em reduce} the energy of its boundary by 
elongating in x-direction. Since the ground state energies of both 
structures are nearly degenerate, the formation of $(2\times1)$ may reduce 
the surface free energy. This picture is confirmed by the fact that islands 
on sublimating surfaces are indeed elongated (See Figure \ref{fig4}a,b) \cite{webpage}. 

\begin{figure}
\begin{center}
\epsfig{file=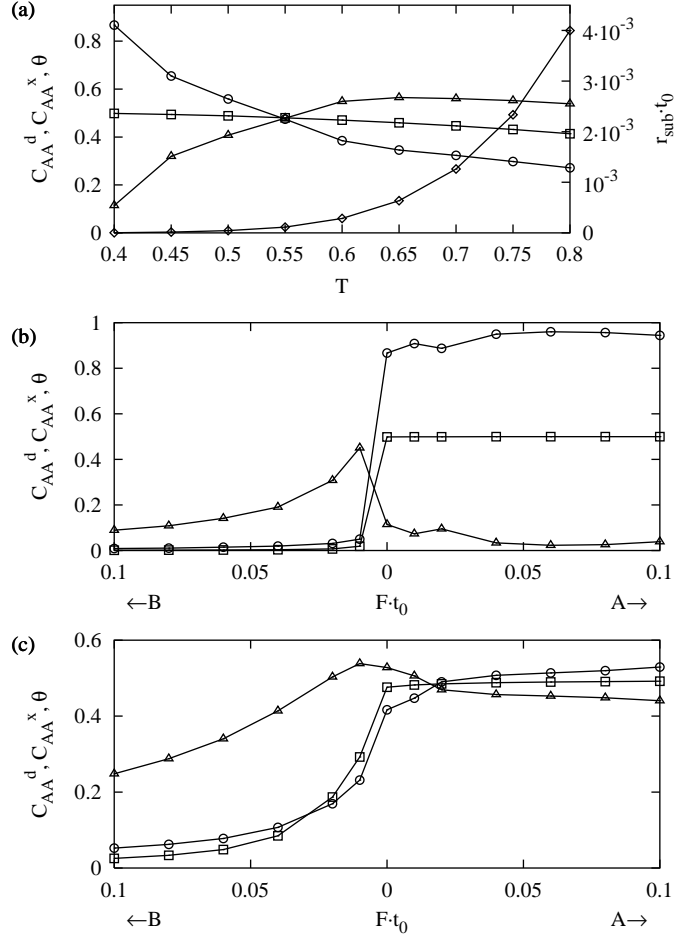,width=0.5\columnwidth,clip=}
\end{center}
\caption{The stationary state of sublimation from vicinal surfaces 
with terraces of $32 \sqrt{2}$ lattice constants width, oriented diagonally. 
(a): Temperature dependence of 
$\theta_{A}$(squares), $C_{AA}^{d}$(circles), $C_{AA}^{x}$ (triangles) and 
$r_{sub}$ (diamonds) under 
vacuum. (b), (c) The influence of an external particle flux on these quantities
at $T= 0.4$ (b) and $T = 0.6$ (c). {\it A} fluxes are shown from left to right, 
{\it B} fluxes from 
right to left. Note that an {\it A}-flux leads to a
re-entry into the $c(2\times2)$ reconstructed configuration at temperatures above 
the reordering in vacuum. Errorbars are on the order of the symbol 
sizes. The system size was $N = 128$. Each datapoint has been obtained in a single 
simulation run.}
\label{fig2}
\end{figure}

\begin{figure}
\begin{center}
\epsfig{file=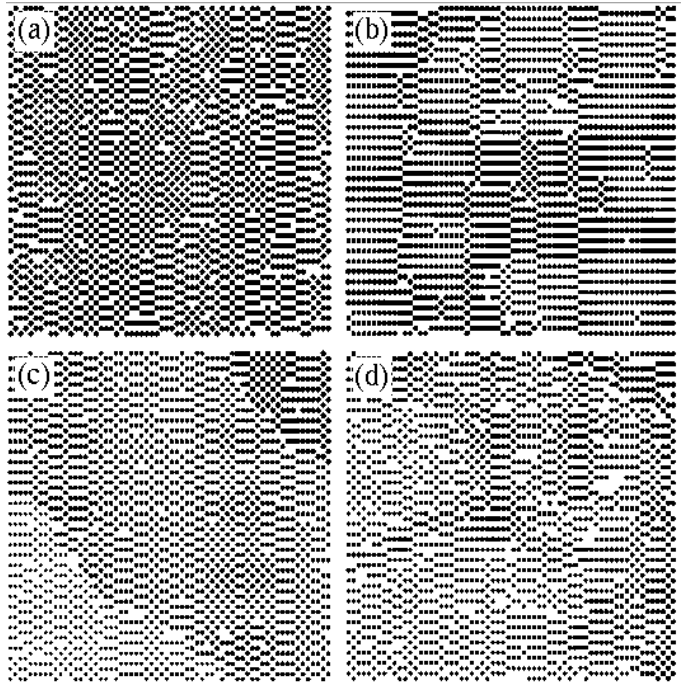,width=0.5\columnwidth,clip=}
\end{center}
\caption{Surface snapshots. The dots show colums terminated 
with {\it A} atoms, while the {\it B} atoms have been omitted for clarity. Larger dots 
denote higher columns.
Panels (a), (b): Layer-by-layer sublimation. Panel (a) shows sections of $64 \times 64$ lattice constans
of the surface in the simulation run of Figure \ref{fig1}a at $t = 7000 t_{0}$ (maximal $I_{spec}$), 
panel (b) at $t = 1.9 \cdot 10^{4} t_{0}$ (minimal $I_{spec}$). 
Panels (c), (d): Sublimation of a vicinal surface in step flow mode. The pictures show sections of 
surfaces at the end of simulation runs shown in 
Figure \ref{fig2}a. Panel (c): $T = 0.5$ (below the 
transition). Panel (d): $T = 0.6$ (above the transition). }
\label{fig4}
\end{figure}

Experiments have revealed a high density of steps on CdTe surfaces
due to the intrinsic surface morphology \cite{nskts00}, which leads to 
a dominant contribution of step flow sublimation. To incorporate this 
effect in our model, we have performed simulations of vicinal surfaces \cite{webpage}. 
We find, that the oszillations in the correlations dissappear for terrace 
widths smaller than $\approx 50$ lattice constants, where step flow 
sublimation dominates. Then, the correlations become stationary apart from 
statistical 
fluctuations.  Figure \ref{fig2}a shows the $T$-dependence of $\theta_{A}$, 
$C_{AA}^{d}$, $C_{AA}^{x}$ and the sublimation rate $r_{sub}$ at a step distance of 
$32\sqrt{2}$ lattice constants. 
The steps were oriented at an angle of 45$^{o}$ to the x-axis, which corresponds to 
the preferential (100)-orientation of steps observed on CdTe(001) \cite{mem99}. 
Snapshots of the surfaces are shown in Figure \ref{fig4}.
Clearly, at $T = 0.55$ there is a transition 
from a $c(2\times2)$ configuration at low temperature to a high-temperature 
regime where the $(2\times1)$ ordering dominates and the material is sublimating.  
This is the conterpart of the phase transition observed in our 
two-dimensional lattice gas model. 
In the investigated temperature 
range $\theta_{A}$ decreases only slightly from $\theta_{A} = 0.498$ at $T = 0.4$ to 
$\theta_{A} = 0.41$ at $T = 0.8$. 

To get insight into the behaviour of a material in an MBE environment, it is 
important to understand how the structure of the surface changes when 
it is exposed to a particle beam. To this end, we have simulated surfaces 
which were exposed to a flux of pure {\it A} or {\it B}. Figure \ref{fig2}b,c 
shows the quantities $\theta_{A}$, $C_{AA}^{d}$ and $C_{AA}^{x}$ as functions 
of the adsorption rate at a temperature $T = 0.4$, which is far below the 
$c(2\times2)$-$(2\times1)$ reordering under vacuum, and at $T = 0.57$, which is
slightly above the transition. 
Applying {\it B} fluxes, we can regulate $\theta_{A}$ to values smaller than those
close to 0.5 observed under vacuum. We find, that a 
decrease of $\theta_{A}$ leads to a significant increase in $C_{AA}^{x}$ (Fig. 
\ref{fig2}b,c). 
This effect is analogous to the higher fraction of $(2\times1)$ ordering on surfaces
with a large number of islands: at low $\theta_{A}$, {\it A}-covered islands form on 
the surface, whose boundaries favour the arrangement of {\it A} atoms in rows. 
On the contrary, an {\it A} flux increases $\theta_{A}$ and $C_{AA}^{d}$ and leads 
to the formation 
of a  dominant $c(2\times2)$ reconstruction even at temperatures above the 
reordering under vacuum. 
The reappearance of a $c(2\times2)$ reconstruction under a Cd-flux at 
high temperature 
and the formation of Te-terminated surfaces with coverage 1 
even at low Te fluxes are well known from experiments on CdTe \cite{ct97}.  
In these experiments however, the formation of Te dimers on the surface 
was observed, an effect 
which is not present in our simple model assuming isotropic interactions 
between {\it B} atoms.

It has been argued \cite{ntss00}, that the CdTe surface is close 
to thermal equilibrium during sublimation, since the bulk serves as a particle
reservoir which defines chemical potentials for both elements. Comparing 
our solid-on-solid model with the two-dimensional anisotropic lattice gas,
we calculate the equilibrium values of 
$C_{AA}^{d}$, $C_{AA}^{x}$ in the planar lattice gas model at the value of $\theta_{A}$ 
we measure for {\it AB} at a given temperature, which can be done using transfer 
matrix techniques \cite{baksv00}. 
Although the non-equilibrium conditions of sublimation  
enhance the dominance of $C_{AA}^{x}$ over $C_{AA}^{d}$ in the high temperature 
regime, we find qualitative agreement in the behaviour of $C_{AA}^{d}$ and $C_{AA}^{x}$.  
However, a simple mapping of the surface layer of the sublimating system on a 2D lattice
gas in thermal equilibrium is not possible. 

Atomic layer epitaxy (ALE) provides a scenario for the investigation 
of our model of the II-VI semiconductor surface in a situation present in technical 
applications. The idea is to obtain self-regulated growth by alternate 
deposition of pure {\it A} and {\it B}. In the absence of reconstructions, 
one would expect the formation of complete monolayers of {\it A} and {\it B} during 
deposition of the elements, yielding growth at a speed of one monolayer (ML)
per cycle. However, this does not necessarily apply in the presence of surfaces 
terminated by vacancy structures with submonolayer coverage. Experimentally, 
in CdTe one finds growth rates of $\approx$ 1 ML/cycle only at $T < 260^{o} C$, 
and growth at a speed of $\approx$ 0.5 ML/cycle at higher temperatures
\cite{dbt96,ct97}.
Figure \ref{fig1}b shows the evolution of $\theta_{A}$, $C_{AA}^{d}$ and 
$C_{AA}^{x}$ during ALE at $T = 0.3$. Each cycle consists of two phases 
of length $t_{cycl} = 4 \cdot 10^{3} t_{0}$. 
In the first phase, 
an {\it A} flux of $5 \cdot 10^{-3} \mbox{ML}/t_{0}$ is applied, 
in the second phase {\it B} is deposited at the same rate. Growth 
was started from a {\it B} terminated flat surface. During the 
first {\it A} phase the surface becomes {\it A}-terminated with $\theta_{A} \approx 1/2$. 
Since $T$ is far below the transition, the reconstruction is $c(2\times2)$. 
At the onset of the {\it B}-phase, $\theta_{A}$ decreases rapidly. This leads to an 
increase of $C_{AA}^{x}$, as reported above for surfaces under a small 
stationary {\it B}-flux. At the end of the cycle, $\approx$50\% of the surface are 
covered with {\it B}-terminated islands, since 
only half of the {\it A} atoms needed for a closed monolayer is present. 
The following {\it A}-phase deposits half a monolayer of {\it A} again. 
However, now the reconstruction is preferentially $(2\times1)$. This is 
analogous to the maximum of $C_{AA}^{x}$ we observe in layer-by-layer 
sublimation after half a monolayer has desorbed, which is due to the 
influence of the island edges on the reconstruction. During the following {\it B} 
phase, {\it A} atoms and {\it AB} pairs on top of the islands diffuse into the gaps 
between the islands, which leads to the formation of a closed {\it B}-terminated 
monolayer.
In our simulations, we observe 2-3 repetitions of this cycle \cite{webpage}. There are two 
effects, which keep it from repeating infinitely. First, we obtain a 
growth rate 
of 0.44 ML/cycle, which is smaller than the ideal value of 0.5 ML/cycle. 
Second, the diffusion of particles from the islands into the gaps is not 
complete. Both effects hinder a perfect closure of the monolayer at the 
end of the even cycles, which leads to a damping of the characteristic 
behaviour. These observations closely resemble the ideas developed to 
explain CdTe growth at a speed of $\approx$0.5 ML/cycle in \cite{dbt96}.

The experimental observation of growth rates of 1 ML/cycle in ALE 
at low temperatures
is inconsistent with our model in its present form, since the hardcore 
repulsion between the {\it A} atoms hinders the deposition of more than 0.5 ML 
of {\it A}. However, this effect might be explained with the existence of 
weakly bound precursor
states, where particles could stay close to the surface until the deposition
of the opposite species allows for an incorporation into the crystal. 
At higher temperatures this reservoir would be inactive due to the weakness 
of the binding. Future research might also address the influence of an 
interlayer diffusion barrier on ALE and MBE in our model.
Additionally, one further step towards a more realistic modelling of CdTe 
should be the simulation of a zinc-blende lattice which, however, requires 
the specification of a greater number of parameters.

In summary, we have presented a simple model of a binary compound material 
which reproduces many effects observed in experiments on CdTe(001) qualitatively. 
We conclude suggesting some experiments to test our model. (1) The 
preparation of CdTe(001) surfaces with an extremely low intrinsic step density
might allow for the observation of oscillations in the correlations during 
layer-by-layer sublimation. 
(2) The deposition of small amounts of 
Te on a CdTe(001) surface should lead to a preferential arrangement of the 
remaining Cd atoms in a $(2\times1)$ ordering. (3) During Cd deposition in 
ALE of CdTe a $c(2\times2)$ reconstruction is expected in cycles starting 
from flat Te-terminated surfaces, while in cycles starting from rough surfaces
$(2\times1)$ should be present.

We thank W. Kinzel and M. Sokolowski for stimulating discussions and a critical 
reading of the manuscript. M.A. was supported by the Deutsche Forschungsgemeinschaft.

\pagebreak 

\end{document}